# Observation of Amplified Stimulated Terahertz Emission from Optically Pumped Epitaxial Graphene Heterostructures

*Taiichi Otsuji[1,3]\*, Hiromi Karasawa[1], Tsuneyoshi Komori[1], Takayuki Watanabe[1],*

*Hirokazu Fukidome[1,3], Maki Suemitsu[1,3], Akira Satou[2,3], and Victor Ryzhii[2,3]*

[1]RIEC Tohoku University, Sendai, Miyagi 980-8577, Japan

[2]University of Aizu, Aizu-Wakamatsu, Fukushima 965-8580, Japan

[3]JST-CREST, Chiyoda-ku, Tokyo 1020075, Japan

\*E-mail: otsuji@riec.tohoku.ac.jp

TITLE RUNNING HEAD    "Coherent, Amplified THz Emission from Graphene Heterostructures"

CORRESPONDING AUTHOR

[1]Email: otsuji@riec.tohoku.ac.jp

Tel./Fax: +81-22-217-6104

Affiliation: Research Institute of Electrical Communication, Tohoku University

Address: 2-1-1 Katahira, Aoba-ku, Sendai, Miyagi 980-8577, Japan

**ABSTRACT**   We experimentally observe the fast relaxation and relatively slow recombination dynamics of photogenerated electrons/holes in an epitaxial graphene-on-Si heterostructure under pumping with a 1550-nm, 80-fs pulsed fiber laser beam and probing with the corresponding terahertz (THz) beam generated by and synchronized with the pumping laser. The time-resolved electric-field intensity originating from the coherent terahertz photon emission is electro-optically sampled in total-reflection geometry. The Fourier spectrum from 1.8 to 5.2 THz agrees well the pumping photon spectrum. This result is attributed to amplified emission of THz radiation from the graphene sample stimulated by the THz probe beam, and provides evidence for the occurrence of negative dynamic conductivity in the terahertz spectral range.

**BRIEFS**   We experimentally observe amplified stimulated emission of terahertz electromagnetic radiation at room temperature from heteroepitaxial graphene on a silicon substrate by optical-pump/terahertz-probe spectroscopy with a femtosecond-pulsed laser operating in the optical communication band.

**MANUSCRIPT**

Graphene, a monolayer of carbon atoms in a honeycomb lattice crystal, has attracted considerable attention due to its unique carrier transport and optical properties, including massless and gapless energy spectra [1-3]. The gapless and linear energy spectra of electrons and holes lead to nontrivial features such as negative dynamic conductivity in the terahertz (THz) spectral range [4], which may lead to the development of a new type of THz laser [5, 6].

To realize such graphene-based devices, understanding the non-equilibrium carrier relaxation/recombination dynamics is critical. Figure 1 presents the carrier relaxation/recombination processes and the non-equilibrium energy distributions of photoelectrons/photoholes in optically pumped graphene at specific times from ~10 fs to picoseconds after pumping. It is known that photoexcited carriers are first cooled and thermalized mainly by intraband relaxation processes on femtosecond to subpicosecond time scales, and then by interband recombination processes. Recently, time-resolved measurements of fast non-equilibrium carrier relaxation dynamics have been carried out for multilayers and monolayers of graphene that were epitaxially grown on SiC [7-11] and exfoliated from highly oriented pyrolytic graphite (HOPG) [12, 13]. Several methods for observing the relaxation processes have been reported. Dawlaty et al. [7] and Sun et al. [8] used an optical-pump/optical-probe technique and George et al. [9] used an optical-pump/THz-probe technique to evaluate the dynamics starting with the main contribution of carrier-carrier (cc) scattering in the first 150 fs, followed by observation of carrier-phonon (cp) scattering on the picosecond time scale. Ultrafast scattering of photoexcited carriers by optical phonons has been theoretically predicted by Ando [14], Suzuura [15] and Rana [16]. Kamprath et al. [12] observed strongly coupled optical phonons in the ultrafast carrier dynamics for a duration of 500 fs by optical-pump/THz-probe spectroscopy. Wang et al. [11] also observed ultrafast carrier relaxation via emissions from hot-optical phonons for a duration of ~500 fs by using an optical-pump/optical-probe technique. The measured optical phonon lifetimes found in these studies were ~7 ps [12], 2-2.5 ps [11], and ~1 ps [9], respectively, some of which agreed fairly well with theoretical calculations by Bonini et al. [17]. A recent study by Breusing et al. [13] more precisely revealed ultrafast carrier dynamics with a time resolution of 10 fs for exfoliated graphene and graphite. It has been shown that the intraband carrier equilibration in optically excited graphene (with pumping photon energy $\hbar\Omega$) first establishes separate quasi-equilibrium distributions of electrons and holes at around the level $\varepsilon_f \pm \hbar\Omega/2$ ($\varepsilon_f$: Fermi energy) within 20-30 fs after excitation (see Fig. 1(b)), followed by cooling of these electrons and holes mainly by emission of a cascade (*N* times) of optical phonons ($\hbar\omega_0$) within 200 fs to occupy the states $\varepsilon_f \pm \varepsilon_N \approx \varepsilon_f \pm \hbar(\Omega/2 - N\omega_0)$, $\varepsilon_N < \hbar\omega_0$ (see Fig. 1(c)). Then, thermalization occurs via electron-hole recombination as well as intraband Fermization due to cc scattering and cp scattering (as shown with energy $\hbar\omega_q$ in Fig. 1(a)) on a picosecond time scale (see Fig. 1(d)), while the interband cc scattering and cp scattering are slowed by the density of states effects and Pauli blocking.

For the electron-hole recombination, radiative recombination via direct-transition to emit photons $\hbar\omega \approx 2\varepsilon_N$ and non-radiative recombination via Auger processes, plasmon emissions, and phonon emissions [9, 18] are considered. In the case of the radiative recombination, due to the relatively small values of $\hbar\omega \approx 2\varepsilon_N < 2\hbar\omega_0$, as well as the gapless symmetrical band structure, photon emissions over a wide THz frequency range are expected if the pumping photon energy is suitably chosen and the pumping intensity is sufficiently high. The incident photon spectra are expected to be reflected in the THz photoemission spectra as evidence that such a process occurs. Previous studies on the observation of carrier recombination dynamics in graphene have been insensitive to the mechanisms of radiative/non-radiative recombination processes, and to date, the observation of such THz photoemissions has not been reported. In this work, we observed amplified stimulated THz emission from optically pumped and THz-probed heteroepitaxial graphene heterostructures and verified the occurrence of negative dynamic conductivity.

The real part of the net ac conductivity Re $\sigma_\omega$ (which is proportional to the absorption coefficient of photons with frequency $\omega$) comprises the contributions of both interband and intraband transitions:

$$\text{Re } \sigma_\omega = \text{Re}\sigma_\omega^{\text{inter}} + \text{Re}\sigma_\omega^{\text{intra}}. \tag{1}$$

The intraband term in Eq. (1) corresponds to the Drude mechanism of THz photon absorption, while the interband term corresponds to the generation/recombination rates under the Dirac-Fermion energy/momentum distribution. Let us consider a practical situation under weak pumping with infrared photons: $T \approx T_0$, $\varepsilon_F < k_B T$, $\hbar\Omega >> k_B T$, $\tau_{rad} >> \tau$, $k_B T \tau / \hbar < \omega^2 \tau^2$, where $T_0$ is the lattice temperature. Under such conditions [4],

$$\text{Re } \sigma_\omega^{\text{inter}} = \frac{e^2}{4\hbar}\left[1 - f_e\left(\frac{\hbar\omega}{2}\right) - f_h\left(\frac{\hbar\omega}{2}\right)\right] \approx \frac{e^2}{8\hbar k_B T}\left(\frac{\hbar\omega}{2} - \varepsilon_F\right) = \frac{e^2}{8\hbar}\left(\frac{\hbar\omega}{2k_B T} - \frac{\pi}{\hbar\Omega}\frac{e^2}{\hbar}\frac{\tau_{rad}}{\Sigma_0}I_\Omega\right), \tag{2}$$

$$\text{Re } \sigma_\omega^{\text{intra}} = \frac{e^2 v_F \tau}{2\pi\hbar^2(1+\omega^2\tau^2)}\int_0^\infty dp\, p\left(-\frac{df}{dp}\right) = \frac{(\ln 2 + \varepsilon_F/2k_B T)e^2}{\pi\hbar}\frac{T\tau}{\hbar(1+\omega^2\tau^2)}, \tag{3}$$

where $T$ is the electron temperature, $e$ is the elementary charge, $\varepsilon_F$ is the Fermi level, $k_B$ is the Boltzmann constant, $\tau_{rad}$ is the electron-hole radiative recombination lifetime, $\tau$ is the intraband momentum relaxation time of electrons/holes, $\hbar$ is the reduced Plank constant, $v_F$ is the Fermi velocity, $\Sigma_0$ is the carrier density in the dark, $f_{(e,h)}$ is the Fermi-Dirac distribution function ($e$: electrons, $h$: holes), $I_\Omega$ is the pumping intensity, and $\omega$ is the angular frequency. The intraband term takes the Drude form, decreasing monotonically and approaching 0 with increasing $\omega$, while the interband term contributes to the linear decrease (increase) in the intraband term in the low (high) $\omega$ region. The break-even frequency depends on $I_\Omega$ and $\tau_{rad}$. Therefore, Re $\sigma_\omega$ is a minimum at a specific frequency: $\omega = \bar{\omega}$.

Around $\omega = \bar{\omega}$, Re $\sigma_\omega$ becomes [4]

$$\text{Re } \sigma_\omega \cong \frac{e^2 \bar{g}}{8\hbar} \left[ 1 + \frac{3}{2}\left(\frac{\omega - \bar{\omega}}{\bar{\omega}}\right)^2 - \frac{I_\Omega}{\bar{I}_\Omega} \right], \tag{4}$$

where $\bar{g} = 2\left(\frac{4\ln 2}{\pi}\right)^{1/3}\left(\frac{\hbar}{k_B T \tau}\right)^{1/3}$, $\bar{\omega} \approx \left(\frac{k_B T \tau}{\hbar}\right)^{2/3}\frac{1.92}{\tau}$, $\bar{I}_\Omega \approx 11\left(\frac{\hbar}{k_B T \tau}\right)^{1/3}\left(\frac{k_B T}{\hbar v_F}\right)^2 \frac{\hbar \Omega}{\tau_{rad}}$, (5)

and $v_F$ is the Fermi velocity ($\sim 10^6$ m/s). As seen from Eq. (2), Re $\sigma_\omega$ is a minimum at $\omega = \bar{\omega}$. When the pumping intensity exceeds the threshold $I_\Omega > \bar{I}_\Omega$, the dynamic conductivity becomes negative. When $T$ = 300 K, $\tau = 10^{-12}$ s, and $\tau_{rad} = 10^{-9} \sim 10^{-11}$ s, this threshold is $\bar{I}_\Omega \approx 60 \sim 6000$ W/cm$^2$. Assuming a device size of 100 μm $\times$ 100 μm we find that the pumping intensity required for negative dynamic conductivity is $\bar{I}_\Omega \approx 6 \sim 600$ mW. The value of 6 mW is a feasible pumping intensity.

When graphene is strongly pumped ($\varepsilon_F > k_B T$), Re $\sigma_\omega$ becomes [4]

$$\mathrm{Re}\,\sigma_\omega \cong \frac{e^2}{2\hbar}\left[-1+0.43(\frac{\bar{\omega}}{\omega})^2\sqrt{\frac{I_\Omega}{\bar{I}_\Omega}}\right]. \tag{6}$$

Equation (6) shows that $\mathrm{Re}\,\sigma_\omega$ changes from negative to positive at extremely high intensity; thus, at $\omega = \bar{\omega}$, for example, $\mathrm{Re}\,\sigma_\omega < 0$ when $\bar{I}_\Omega < I_0 < 5\bar{I}_\Omega$ [4].

In order to verify the proposed concept, we conduct an experimental study on the electromagnetic radiation emitted from an optically pumped graphene structure. The sample used in this experiment is heteroepitaxial graphene film grown on a SiC(110) thin film heteroepitaxially grown on a Si(110) substrate via thermal graphitization of the SiC surface [19-21]. In the Raman spectrum of the graphene film, the principal bands of graphene, namely, the G (1595 cm$^{-1}$) and G' (2730 cm$^{-1}$) bands, are observed. Furthermore, transmission electron microscopy images indicate that the film is stratified. It is thus concluded that epitaxial graphene with a planar structure can be produced by this fabrication method. Furthermore, the epitaxial graphene layer is inferred to have a non-Bernal stacking arrangement because the G' band in the Raman spectrum can be expressed as a single component related to the two-dimensionality of the graphene film [22, 23]. The non-Bernal stacked epitaxial graphene layers grown by our method can be treated as a set of isolated single graphene layers, as in the case of an epitaxial graphene layer on a C-terminated SiC bulk crystal [24]. The G-band peak at 1595 cm$^{-1}$ corresponds to an optical phonon energy at the zone center of 197.8 meV.

We measure the carrier relaxation and recombination dynamics in optically pumped epitaxial graphene-on-silicon (GOS) heterostructures using THz time-domain spectroscopy based on an optical pump/THz-and-optical-probe technique. The time-resolved field emission properties are measured by an electro-optic sampling method in total-reflection geometry [25]. To obtain the THz photon emissions from the above-mentioned carrier relaxation/recombination dynamics, the pumping photon energy (wavelength) is carefully selected to be around 800 meV (1550 nm). To perform intense pumping beyond the threshold, a femtosecond pulsed fiber laser with full width at half-maximum (FWHM) of 80 fs, pulse energy of 50 pJ/pulse, and frequency of 20 MHz was used as the pumping source. The setup is shown in Fig. 2. The graphene sample is placed on the stage and a 100-μm-thick (101)-oriented CdTe crystal is placed onto the sample; the CdTe crystal acts as a THz probe pulse emitter as well as an electro-optic sensor. The single femtosecond fiber laser beam is split into two beams: one for optical pumping and generating the THz probe beam, and one for optical probing. The pumping laser, which is linearly polarized and mechanically chopped at ~1.2 KHz, is simultaneously focused at normal incidence onto the sample and the CdTe from below, while the probing laser, which is cross-polarized to the pumping beam, is focused from above. The incident pumping beam is defocused on the sample to satisfy the above-mentioned pumping power requirements. The resulting photoexcited carrier density is $\sim 8 \times 10^{10}$ cm$^{-2}$, which is comparable to the background carrier density. Owing to second-order nonlinear optical effects, the CdTe crystal can rectify the pumping laser pulse to emit THz envelope radiation. If the pumping laser polarization is aligned to be Raman-active for CdTe, the third-order nonlinear effects of the CdTe excite the phonon-polariton to emit THz oscillatory radiation. These THz pulses irradiate the graphene sample, acting as THz probe signals to stimulate THz photon emission via electron-hole recombination in the GOS. Due to the geometrical situation of the experimental setup as shown in Fig. 2, the time delay of the THz probe with respect to optical pumping is fixed at around 200-300 fs so that the photoelectrons/holes are stimulated immediately after losing their energy via the cascade of optical phonon emissions, when they still have a distribution of $\varepsilon_f \pm \varepsilon_N$ (as is shown in Fig. 1(c)). On the other hand, through the Si prism attached to the CdTe crystal, the optical probing beam is totally reflected back to the lock-in detection block, and its phase information reflecting the electric field intensity is lock-in amplified. By sweeping the timing of the optical probe using an optical delay line, the whole temporal profile of the field emission properties can be obtained. The system bandwidth is estimated to be around 6 THz, which is limited mainly by the Reststrahlen band of the CdTe sensor crystal.

Figure 3 shows the autocorrelations and spectral profiles of the pumping laser beams under two different pulse compression conditions. The horizontal axis indicates the wavelength and frequency together with the estimated THz photon frequency to be emitted from the sample. The dotted line plots the dynamic conductivity at a pumping intensity twice as high as the threshold intensity calculated for 300 K using Eqs. (1)-(3) with a $\tau$ value of $1\times10^{-12}$ ps. The shaded area shows the negative dynamic conductivity. First, the experiment was conducted with the pumping pulse shown in Fig. 3 as red lines.

Figure 4 shows the measured temporal response (inset) and the corresponding Fourier spectrum (solid lines). The dashed line in Fig. 4 is the photoemission spectrum predicted from the pumping laser spectrum. The emission spectrum from CdTe without graphene shows a dominant peak around 5 THz, and a weak side lobe extends to below 1 THz (blue line in Fig. 4). On the other hand, the results with graphene agree well with the THz photon spectrum predicted from the pumping photon spectrum. The results include an additional peak around 5 THz (red line in Fig. 4). It is thought that the THz emissions from graphene are stimulated by the coherent THz probe radiation that originates from the phonon-polaritons (LO phonon at 5.1 THz, TO phonon at 4.3 THz, and soft-TO phonon at 2.1 THz) in Raman-active CdTe excited by the pump laser beam. The THz emissions are amplified by photoelectron/hole recombination in the range of the negative dynamic conductivity.

Figure 5(a) shows another case with a different THz probe spectrum. The CdTe crystal is placed in the axial direction, becoming Raman inactive. In this case, the emission from the CdTe without graphene exhibits a temporal response similar to optical rectification with a single peak at around 1 THz and an upper weak side lobe extending to around 7 THz (blue lines in Fig. 5). On the other hand, as in the case of Fig. 4, the results with graphene agree well with the pumping photon spectrum and include an additional peak around 1 THz from the original CdTe spectrum (red lines in Fig. 5). In order to confirm with certainty that the emissions from graphene reflect the pumping photon spectrum, the pulse compression of the pumping laser pulse was altered to broaden its spectrum as shown by the black lines in Fig. 3. Due to the pulse compression condition, the pulse energy increases ~100 pJ/pulse. As can be seen in Fig. 5(b), the emission spectrum of the graphene was broadened, corresponding to the pumping photon energy.

Furthermore, to confirm the effects of the THz probe, we replace the first CdTe crystal with another CdTe crystal having a high-reflectivity coating for IR on its bottom surface, in order to eliminate generation of the THz probe signal. In this case, no distinctive response is observed with or without graphene. Since the measurements are taken as an average, the observed response is undoubtedly a coherent process that cannot be obtained via spontaneous emission processes, providing clear evidence of stimulated emission. For all three cases of emissions from graphene in Figs. 4 and 5, the lower cutoff around 2 THz (8.3 meV) is slightly higher than the theoretical estimation (~5 meV) [5]. This may be due to a small bandgap forming due to the interlayer coupling of the existing multilayer graphene or substrate-induced asymmetric potential deformation [26].

From the above results, it is inferred that THz emissions from graphene are stimulated by the coherent THz probe radiation. Furthermore, the THz emissions are amplified via photoelectron/hole recombination in the range of the negative dynamic conductivity. In conclusion, we have successfully observed coherent amplified stimulated THz emissions arising from the fast relaxation and relatively slow recombination dynamics of photogenerated electrons/holes in an epitaxial graphene heterostructure. The results provide evidence of the occurrence of negative dynamic conductivity, which can potentially be applied to a new type of THz laser.

ACKNOWLEDGMENTS

This work was financially supported in part by the JST-CREST program, Japan, and a Grant-in-Aid for Basic Research (S) from the Japan Society for the Promotion of Science.

FIGURE CAPTIONS

**Figure 1.** Schematic view of graphene band structure (a) and energy distributions of photogenerated electrons and holes (b)-(d). Arrows denote transitions corresponding to optical excitation by photons with energy $\hbar\Omega$, cascade emission of optical phonons with energy $\hbar\omega_0$, and radiative recombination with emission of photons with energy $\hbar\omega$. (b) after ~20 fs from optical pumping, (c) after ~200 ps from optical pumping, (d) after ~1 ps from optical pumping.

**Figure 2.** Measurement setup for optical-pump/THz-and-optical-probe spectroscopy. CdTe crystal on top of the graphene sample generates THz probe pulse and allows electro-optic detection of THz electric field intensity.

**Figure 3.** Temporal (autocorrelation) and spectral profiles of pumping laser beam for two pulse compression conditions used in this experiment. Dotted line denotes the dynamic conductivity normalized to the characteristic conductivity $e^2/2\hbar$ at a pumping intensity of twice the threshold intensity at 300 K calculated using Eqs. (1)-(3). Shaded area denotes negative dynamic conductivity.

**Figure 4.** Measured field emission properties (inset: temporal responses; main plot: Fourier spectrum) when the THz probe beam is generated by excitation of coherent phonon-polaritons in CdTe. Dashed line is the photoemission spectrum predicted from the pumping laser spectrum.

**Figure 5.** Measured field emission properties (inset: temporal responses; main plot: Fourier spectrum) when the THz prove beam is generated by optical rectification of pumping photons in CdTe. Dashed

line is the photoemission spectrum predicted from the pumping laser spectrum. (a) Pumping with pulses similar to those used for spectra in Fig. 4, and (b) pumping with shorter pulses (broader spectral).

REFERENCES

[1] Geim, K.; and Novoselov, K. S.; "The rise of graphene," *Nat. Mater.* **2007**, 6, 183.

[2] Novoselov, K.S.; Geim, A.K.; Morozov, S.V.; Jiang, D.; Katsnelson, M. I.; Grigorieva, I.V.; Dubonos, S.V.; Firsov, A.; "Two-dimensional gas of massless Dirac fermions in graphene," *Nature* **2005**, 438, 197.

[3] Kim, P.; Zhang, Y.; Tan, Y.-W.; Stormer, H. L.; "Experimental observation of the quantum Hall effect and Berry's phase in graphene," *Nature* **2005**, 438, 201.

[4] Ryzhii,V.; Ryzhii, M.; Otsuji, T.; "Negative dynamic conductivity of graphene with optical pumping," *J. Appl. Phys.* **2007**, 101, 083114.

[5] Dubinov, A. A.; Aleshkin, V. Y.; Ryzhii, M.; Otsuji, T.; and Ryzhii, V.; "Terahertz laser with optically pumped graphene layers and Fabri–Perot resonator," *Appl. Phys. Express* **2009**, 2, 092301.

[6] Ryzhii, V.; Ryzhii, M.; Satou, A.; Otsuji, T.; Dubinov, A. A.; and Aleshkin, V. Y.; "Feasibility of terahertz lasing in optically pumped epitaxial multiple graphene layer structures," *J. Appl. Phys.* **2009**, 106, 084507.

[7] Dawlaty, J.M.; Shivaraman, S.; Chandrashekhar, M.; Rana, F.; Spencer, M. G.; "Measurement of ultrafast carrier dynamics in epitaxial graphene," *Appl. Phys. Lett.* **2008**, 92, 042116.

[8] Sun, D.; Wu, Z.-K.; Divin, C.; Li, X.; Berger, C.; de Heer, W. A.; First, P.N.; Norris, T. B.; "Ultrafast Relaxation of Excited Dirac Fermions in Epitaxial Graphene Using Optical Differential Transmission Spectroscopy," *Phys. Rev. Lett.* **2008**, 101, 157402.

[9] George, P.A.; Strait, J.; Dawlaty, J.; Shivaraman, S.; Chandrashekhar, M.; Rana, F.; Spencer, M.G.; "Ultrafast Optical-Pump Terahertz-Probe Spectroscopy of the Carrier Relaxation and Recombination Dynamics in Epitaxial Graphene," *Nano Lett.* **2008**, 8, 4248.

[10] Choi, H.; Borondics, F.; Siegel, D. A.; Zhou, S. Y.; Martin, M. C.; Lanzara, A.; Kaindl, R.A.; "Broadband electromagnetic response and ultrafast dynamics of few-layer epitaxial graphene," *Appl. Phys. Lett.* **2009**, 94, 172102.

[11] Wang, H.; Strait, J.H.; George, P.A.; Shivaraman, S.; Shields, V.B.; Chandrashekhar, M.; Hwang, J.; Rana, F.; Spencer, M.G.; Ruiz-Vargas, C.S.; and Park, J.; "Ultrafast relaxation dynamics of hot optical phonons in Graphene," *Arxiv* **2009**, 0909.4912.

[12] Kampfrath, T.; Perfetti, L.; Schapper, F.; Frischkorn, C.; and Wolf, M.; "Strongly coupled optical phonons in the ultrafast dynamics of the electronic energy and current relaxation in graphite," *Phys. Rev. Lett.* **2005**, 95, 187403.

[13] Breusing, M.; Ropers, C.; and Elsaesser, T.; "Ultrafast carrier dynamics in graphite," *Phys. Rev. Lett.* **2009**, 102, 086809.

[14] Ando, T.; "Anomaly of optical phonon in monolayer graphene," *J. Phys. Soc. Jpn.* **2006**, 75, 124701.

[15] Suzuura, H.; Ando, T.; "Zone-boundary phonon in graphene and nanotube," *J. Phys. Soc. Jpn.* **2008**, 77, 044703.

[16] Rana, F.; George, P.A.; Strait, J.H.; Dawlaty, J.; Shivaraman, S.; Chandrashekhar, M.; and Spencer, M.G.; "Carrier recombination and generation rates for intravalley and intervalley phonon scattering in graphene," *Phys. Rev. B* **2009**, 79, 115447.

[17] Bonini, N.; Lazzeri, M.; Marzari, N.; Mauri, F.; "Phonon anharmonicities in graphite and graphene," *Phys. Rev. Lett.* 2007, 99, 176802.

[18] Rana, F.; "Electron-hole generation and recombination rates for Coulomb scattering in graphene," *Phys. Rev. B* **2007**, 76, 155431.

[19] Suemitsu, M.; Miyamoto, Y.; Handa, H.; Konno, A.; "Graphene formation on a 3C-SiC(111) thin film grown on Si(110) substrate, " *e-J. Surface Sci. Nanotech.* **2009**, 7, 311.

[20] Miyamoto, Y.; Handa, H.; Saito, E.; Konno, A.; Narita, Y.; Suemitsu, M.; Fukidome, H.; Ito, T.; Yasui, K.; Nakazawa, H.; Endoh, T.; "Raman-Scattering Spectroscopy of Epitaxial Graphene Formed on SiC Film on Si Substrate," *e-J. Surface Sci. Nanotech.* **2009**, 7, 107.

[21] Fukidome, H.; Miyamoto, Y.; Handa, H.; Saito, E.; Suemitsu, M.; "Epitaxial Growth Processes of Graphene on Silicon Substrates," *Jpn. J. Appl. Phys.*, in press.

[22] Cançadoa, L.G.; Takaia, K.; Enokia, T.; Endob, M.; Kimb, Y.A.; Mizusakib, H.; Spezialic, N.L.; Jorioc A.; and Pimentac, M.A.; "Measuring the degree of stacking order in graphite by Raman spectroscopy," *Carbon* **2008**, 46, 272.

[23] Faugeras, C.; Nerrière, A.; Potemski, M.; Mahmood, A.; Dujardin, E.; Berger, C.; and de Heer, W.A.; "Few-layer graphene on SiC, pyrolitic graphite, and graphene: A Raman scattering study," *Appl. Phys. Lett.* **2008**, 92, 011914.

[24] Hass, J.; Varchon, F.; Millán-Otoya, J. E.; Sprinkle, M.; Sharma, N.; de Heer, W. A.; Berger, C.; First, P. N.; Magaud, L.; and Conrad, E. H.; "Why multilayer graphene on 4H-SiC(000$\bar{1}$) behaves like a single sheet of graphene," *Phys. Rev. Lett.* **2008**, 100, 125504.

[25] Min, L.; Miller, R.J.D.; "Sub-picosecond reflective electro-optic sampling of electron-hole vertical transport in surface-space-charge field," *Appl. Phys. Lett.* **1990**, 56, 524.

[26] Sano, E.; Otsuji, T.; "Theoretical evaluation of channel structure in graphene field-effect transistors ," *Jpn. J. Appl. Phys.* **2009**, 48, 041202.

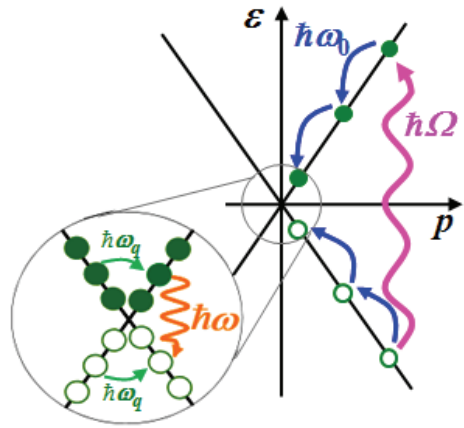

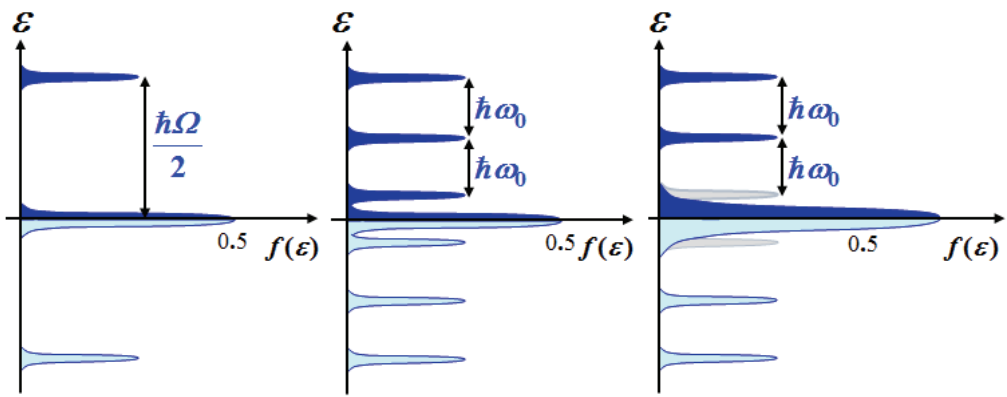

(a)     (b)     (c)     (d)

Figure 1.

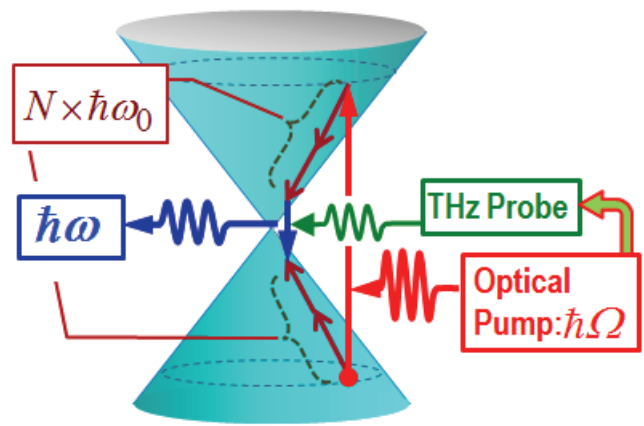

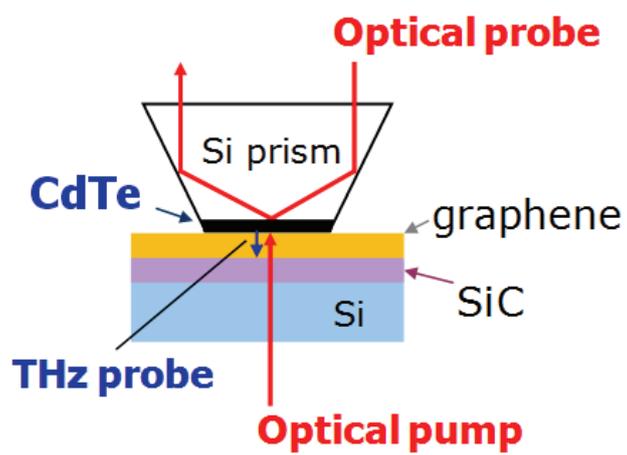

Figure 2.

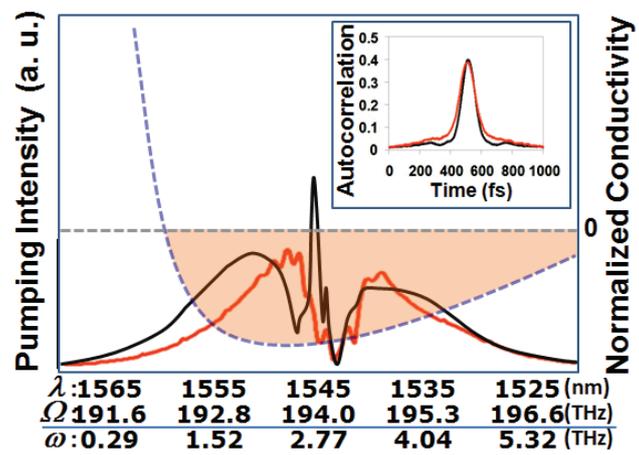

Figure 3.

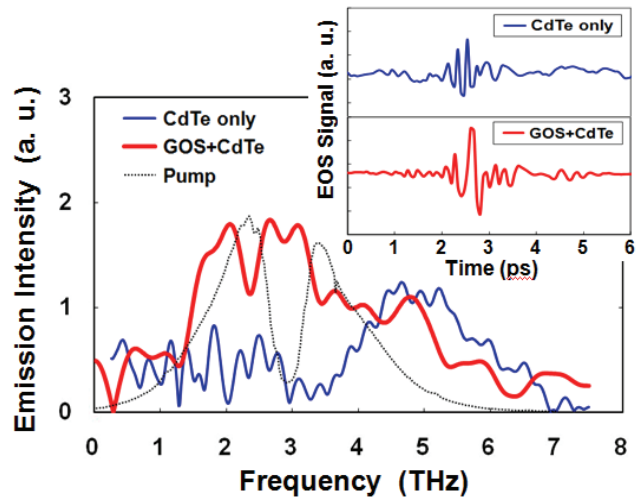

Figure 4.

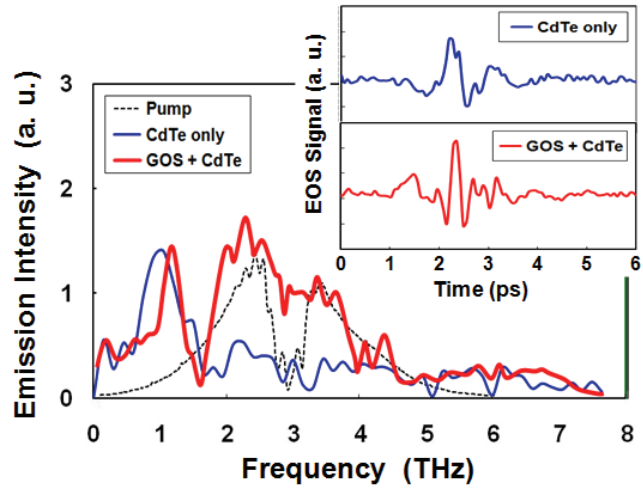

(a)

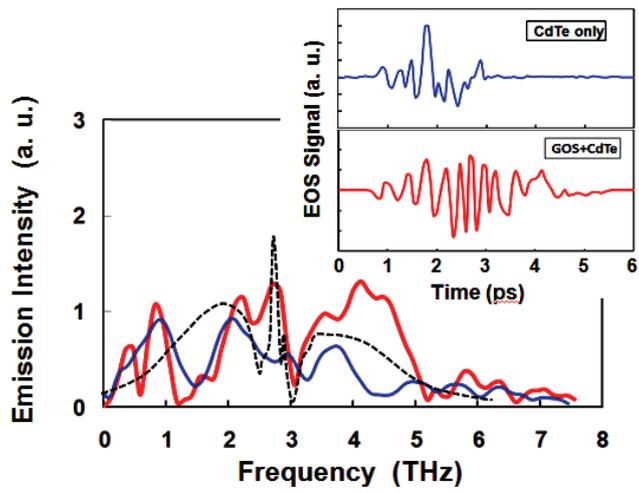

(b)

Figure 5.

**Supporting Information Available**

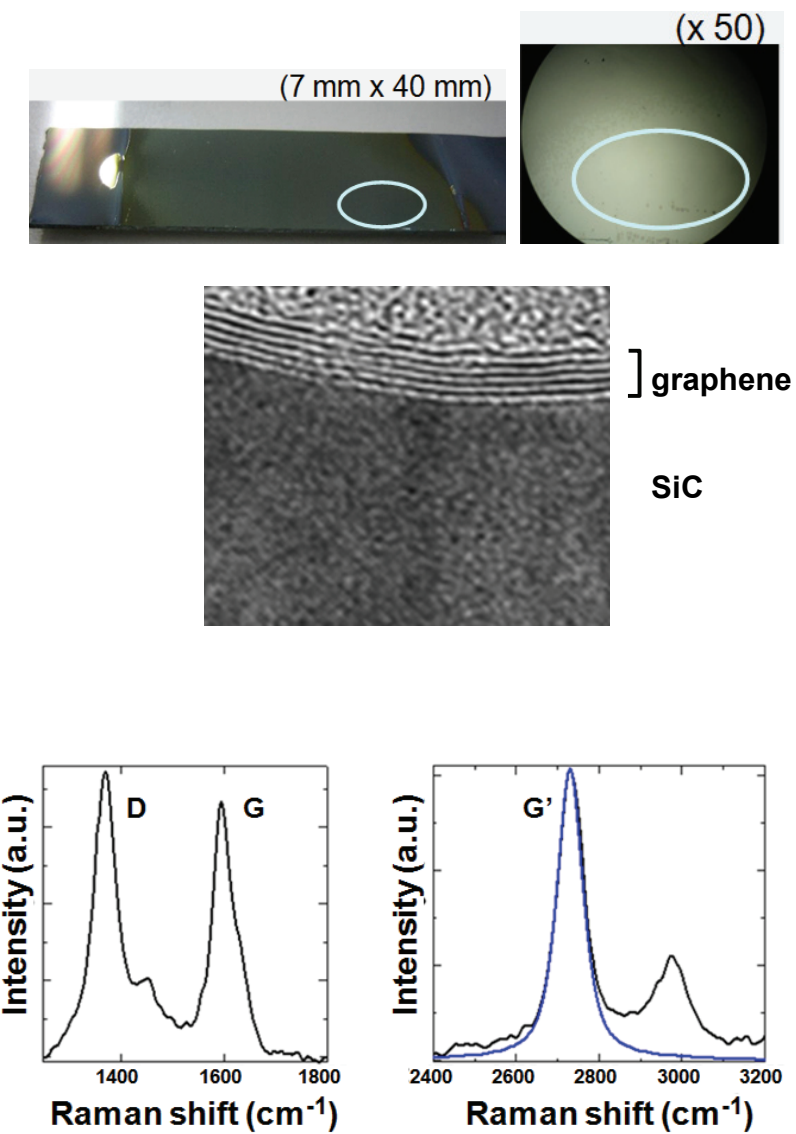

Here provides Figures showing a heteroepitaxial graphene on silicon sample. Upper: photo, micro-zoom images, middle: TEM image, bottom: Raman spectra at D, G, and G' band.